\documentstyle[12pt]{article}
\begin{document}
\hfill SINP/TNP/94-05\\
\smallskip
\hfill hep-th/9405032\\
\begin{center}
{\Large{\bf Fermion scattering off
dilatonic black holes}}\\[1cm]
{\large A. Ghosh and P. Mitra}\footnote{
e-mail mitra@saha.ernet.in}\\[.4cm]
Saha Institute of Nuclear Physics\\
Block AF, Bidhannagar\\
Calcutta 700 064, INDIA\\
\end{center}

\bigskip
\abstract{  The  scattering of massless fermions off magnetically
charged dilatonic  black holes is reconsidered and a violation of
unitarity is  found.  Even  for a single species of fermion it is
possible for a particle to disappear into the black hole with its
information content.}

\vspace{1cm}
In  recent  years there has been a lot of interest in the physics
of black holes. The issue which has engaged the attention of most
workers is that of possible information loss. Matter falling into
a black hole carries  some  information  with  it.  This  becomes
inaccessible  to  the  rest of the world, but may in principle be
supposed to be stored inside the black  hole  in  some  sense.  A
problem arises when the black hole evaporates through the process
of  Hawking  radiation.  The information does seem to be lost now
\cite{Hawk}.

Although there have been attempts at studying this problem in its
full   complexity   \cite{tH}, most authors have considered
simplified models of black holes as in \cite{Garf, CGHS};
see \cite{HS} for a review.
We shall consider the extremal magnetically  charged  black  hole  solution
of dilatonic gravity. This is
a  four dimensional model involving an extra field -- the dilaton
-- but for s-wave scattering of particles in the  field  of  this
black  hole,  the  angular coordinates are not relevant and a two
dimensional  effective  action  can  be  used \cite{GSB}.  If  the  energies
involved  are  not too high, the metric and the dilaton field can
be treated  as  external  classical  quantities  and  an  amusing
version  of  electrodynamics emerges, where the kinetic energy of
the  gauge   field   has   a   position   dependent   coefficient
\cite{Strom}.

The  scattering  of massless fermions has been considered in this
context. The model admits a solution which is very close  to  the
conventional  solution  of  the  Schwinger model, {\it i.e.}, two
dimensional massless electrodynamics. In this solution, there  is
a massive free particle, but in the present case its mass becomes
position  dependent  \cite{Strom, PST}.  To be precise, the mass
vanishes  near  the  {\it mouth}  of  the  black  hole  but   increases
indefinitely as one goes into the {\it throat}. (The dilatonic field
increases linearly with distance in the throat.) This is interpreted to
mean that massless fermions proceeding into the black hole cannot
go very far and have to turn back with probability one. Thus  the
danger of information loss is averted very simply.

In  this  note, we reexamine the model by taking into account the
possibility   of   alternative  solutions.  The  Schwinger  model
possesses other solutions besides the conventional one,  although
this  is  not very well realized by everyone. These correspond to
different quantum theories built from the same classical  theory.
Different  quantum theories can  be
constructed    without    violating     gauge     invariance   by
changing  the  definition  of  the point split fermionic currents
\cite{Hag, BGM}. By
considering this  freedom,  we  shall  demonstrate
that  the  problem of information loss can in fact appear even in
the extremal magnetically charged black hole in a dilatonic background.

The   model  is  described  by  the  Lagrangian density
\cite{Strom, PST}
\begin{equation}
{\cal   L}=\overline{\psi}(i\partial\!\!\!/+eA\!\!\!/)\psi
-{1\over  4}e^{-2\varphi(x)}F^{\mu\nu}F_{\mu\nu},\label{1}
\end{equation}
where  the Lorentz indices take the values 0,1 corresponding to a
(1+1)$-$ dimensional spacetime, $e$ measures the coupling of  the
vector  current  corresponding  to the massless fermion $\psi$ to
the  gauge  field  $A$,  and  there  is  a  dilatonic  background
$\varphi(x)$  whose  dynamics we do not go into. It is clear that
if $\varphi(x)$ vanishes, we get the well- known Schwinger  model
\cite{Sch, LS, CKS}. The model with nonvanishing $\varphi(x)$ has
also  been  solved  \cite{Strom, PST} with the help of the usual
scheme of bosonization. Here we discuss a solution in a different
framework, which leads to vastly altered physical consequences.

In two dimensions we can always set
\begin{equation}
A_\mu  =  -{\sqrt{\pi}\over   e}(\widetilde\partial_\mu\sigma   +
\partial_\mu\widetilde\eta),\label{2}
\end{equation}
where \begin{equation}
\widetilde\partial_\mu = \epsilon_{\mu\nu}\partial^\nu,\end{equation}
with $\epsilon_{01}=+1$ and
$\sigma,\widetilde\eta$  are scalar fields.

We shall restrict ourselves to the Lorentz gauge.
{}From (\ref{2}) we see that the field  $\widetilde\eta$  can
be  taken  as  a  massless  field  with $\Box\widetilde\eta=0$. We
introduce its dual through
\begin{equation}
\widetilde\partial_\mu\eta(x)=\partial_\mu\widetilde\eta(x).\end{equation}
These  massless  fields  have  to  be  regularized
\cite{LS} but we  shall not need the explicit form of the
regularization.

The Dirac equation in the presence of the gauge field is
\begin{equation}
\left[i\partial \!\!\!/ + eA\!\!\!/\right]\psi(x)=0.
\end{equation}
This equation is satisfied by
\begin{equation}
\psi    (x)=:~e^{i\sqrt{\pi}\gamma_5   \left[\sigma   (x)   +\eta
(x)\right]}:\psi^{(0)}(x),\label{7}
\end{equation}
where, $\psi^{(0)}(x)$ is a free fermion field
satisfying $i{\partial\!\!\!/}\psi^{(0)}(x)=0$.

We can calculate  the  gauge  invariant current using the point -
splitting regularization. While constructing  a  gauge  invariant
bilinear  of fermions which in the limit of zero separation would
give the  usual  fermion  current,  we  can generalize  the
conventional construction \cite{Sch}. We take
\begin{eqnarray}
J_\mu^{reg}(x)&=&\lim_{\epsilon\rightarrow 0}[\overline\psi(x+\epsilon)
\gamma_\mu~:e^{ie\int^{x+\epsilon}_x dy^\rho\big\{ A_\rho(y)-2
\partial^\nu[\Phi(y) F_{\rho\nu}(y)]\big\} }:\psi(x)\nonumber\\
&-&  {\rm v.e.v.}]
\end{eqnarray}
where    $\Phi(y)$     is     a  nondynamical function  of
spacetime coordinates
which we shall fix later on. The addition of a term
containing   this   function in  the  exponent represents   a
generalization   of   Schwinger's   regularizing   phase   factor
\cite{Hag, BGM}. It
preserves gauge invariance, Lorentz invariance and even
the  linearity  of the theory. The explicit coordinate
dependence  of  $\Phi$  may  come  as  a surprise, but it must be
remembered that the  model  under  discussion  does  not  possess
translation   invariance   because   of   the  factor  containing
$\varphi(x)$ in the Lagrangian density (\ref{1}).
In fact this freedom can be used to simplify the solution
of the model enormously, as we shall see.
Now using (\ref{2}) and (\ref{7}) together with
\begin{equation}
F_{\mu\nu}={\sqrt\pi\over e}\epsilon_{\mu\nu}\Box\sigma\label{F}
\end{equation}
we obtain the current  which,  upto an overall
wavefunction renormalization,  is equal to
\begin{eqnarray}
J_\mu^{reg}(x) &\approx & ~:\overline{\psi^{(0)}}(x)\gamma_\mu\psi^{(0)}(x):-
i\sqrt\pi\lim_{\epsilon\rightarrow 0}\langle 0\mid
\overline{\psi^{(0)}}(x+\epsilon)\gamma_\mu
\bigl[(\gamma_5\epsilon \cdot\partial
\nonumber\\& +&\epsilon\cdot\widetilde\partial)(\sigma+\eta)
+2\epsilon\cdot\widetilde\partial(\Phi\Box\sigma)\big]
\psi^{(0)}(x)\mid 0\rangle\\
&=& ~:\overline{\psi^{(0)}}(x)\gamma_\mu\psi^{(0)}(x):\nonumber\\&-&
{1\over \sqrt \pi}\big[{\epsilon_\mu
\epsilon_\nu-\widetilde\epsilon_\mu\widetilde\epsilon_\nu\over
{\epsilon^2}}\widetilde\partial^\nu(\sigma+\eta)+ 2{
\epsilon_\mu\epsilon_\nu\over {\epsilon^2}}\widetilde\partial^\nu
(\Phi\Box\sigma)\big],
\end{eqnarray}
where we have used the identity
\begin{equation}
\langle 0\mid\overline{\psi^{(0)}}_\alpha  (x+\epsilon)\psi_\beta (x)\mid
0\rangle =-i{\epsilon\!\!\!/_{\beta\alpha}\over 2\pi\epsilon^2}.
\end{equation}
Now we take  the  symmetric limit {\it i.e.\/} average over the
point splitting directions $\epsilon$ and finally obtain
\begin{equation}
J_\mu^{reg}(x)=-{1\over \sqrt\pi}\widetilde\partial_\mu(\phi+\sigma+\Phi\Box
\sigma+\eta),
\end{equation}
where $\phi$ is a free massless bosonic field satisfying
\begin{equation}
-{1\over{\sqrt\pi}}\widetilde\partial_\mu\phi=:\overline\psi^
{(0)}(x)\gamma_\mu\psi^{(0)}(x):\end{equation}   and  thus  representing  the
conventional bosonic equivalent of the free
fermionic   field   $\psi^{(0)}$
\cite{Cole}.
We find
\begin{eqnarray}
J_{\mu 5}^{reg}(x)&=&\epsilon_{\mu\nu}J^\nu_{reg}(x)\\
&=&-{1\over{\sqrt\pi}}
\partial_\mu(\phi +\eta +
\sigma +\Phi\Box\sigma).
\end{eqnarray}
This implies that the anomaly in this regularization is
\begin{equation}
\partial^\mu  J_{\mu 5}^{reg}=
-{1\over{\sqrt\pi}}
\Box(\phi +\eta +
\sigma +\Phi\Box\sigma).\label{j}\end{equation}

Note now that Maxwell's equation with sources, {\it viz.},
\begin{equation}
\partial_\nu\left({ F^{\nu\mu}\over g^2}\right)
+ eJ^\mu_{reg} =0,
\end{equation}
where
\begin{equation} g^2(x)=e^{2\varphi(x)},
\end{equation}
can be converted to the pair of equations
\begin{equation}
\left[~\Big({1\over g^2}+{e^2\over \pi}
\Phi\Big)\Box~+~{e^2\over \pi}~\right]\sigma=0\label{I}
\end{equation}
and \begin{equation}
\phi +\eta=0.\label{sub}\end{equation}

The  first  equation  (\ref{I}),  which  depends on the choice of
$\Phi$, determines the spectrum of particles in the theory.
The  other  equation  (\ref{sub}), relating two massless free
fields, has to be  satisfied  in  a  weak  sense  by  imposing  a
subsidiary  condition  \begin{equation}  (\phi  + \eta)^{(+)}\mid
phys\rangle =0\end{equation} to select out a physical subspace of
states. One can ensure that $\phi +  \eta$  creates  only  states
with  zero  norm  by taking $\eta$ to be a negative metric field,
{\it i.e.}, by taking its commutators to have the ``wrong'' sign.
The subsidiary condition  then  separates  out  a  subspace  with
nonnegative metric as usual.

$\Phi$ is as yet undetermined. We shall consider a  few  possible
choices.  The  conventional  choice  \cite{Strom,  PST}  is zero.
(\ref{I}) then becomes
\begin{equation}
\left[
\Box~+~{e^2g^2\over \pi}~\right]\sigma=0.
\end{equation}
This describes a particle of mass ${eg(x)\over\sqrt\pi}$. Now
$g$ is related to $\varphi$, which is taken to vary linearly with
distance in the throat of the black hole. The situation envisaged
is that $g$ vanishes at the mouth of the black  hole,  but  rises
indefinitely  as  one  proceeds  into the interior. The effect is
that the mass of the particle vanishes at  the  mouth  but  rises
indefinitely  inside  the  throat.  Since  massless  scalars  are
equivalent to massless fermions in  two  dimensions,  it  follows
that  one  can  think  of  an  initial condition where a massless
fermion starts at the  mouth  of  the  black  hole  and  proceeds
inwards.  The  fact  that  the  mass  involved in the equation of
motion rises  indefinitely  means  that  the  fermion  cannot  go
arbitrarily far and is reflected back with unit probability. Thus
the scattering of the fermion off the black hole is  unitary  and
information is not lost.

On the other hand, if $\Phi$ is chosen to satisfy the condition
\begin{equation} {e^2\over\pi}\Phi= g^2,
\end{equation}
(\ref{I}) simplifies to
\begin{equation}
\left[\Box~+~{e^2\over \pi(g^2+g^{-2})}~\right]\sigma=0.
\end{equation}
The  mass of the particle now vanishes not only at the mouth, but
also asymptotically in the interior of the black hole.  In  fact,
the  mass  has  a  maximum  somewhere in between. Therefore it is
possible for a massless fermion to exist both at the mouth and in
the  interior,  and the height of the barrier being finite, there
is a finite amplitude for the fermion to go in and get lost. Thus
the danger of information loss is {\it not} averted in this case.

A somewhat mundane case is when $\Phi$ is such that
\begin{equation} \left({1\over g^2}+{e^2\over\pi}\Phi\right)=1,
\end{equation}
and (\ref{I}) simplifies to
\begin{equation}
\left[\Box~+~{e^2\over \pi}~\right]\sigma=0.
\end{equation}
This means that the  usual  massive  free  scalar  field  of  the
Schwinger   model   is   recovered.  The modified Schwinger model
thus  accommodates  the  unmodified  solution  with  this altered
definition of currents.

A more dramatic case is when $\Phi$ is allowed to go to infinity.
In  this  case the free scalar field becomes exactly massless. So
the fermion is massless at all positions.  This  fermion  travels
freely into the black hole and {\it all} information is lost.

To  summarize,   we  have  looked  at  the  extremal magnetically
charged black hole in a dilatonic background
using a  generalized  construction of fermion bilinears.  We
point  -  split  the  current  which  is  formally defined as the
product of two fermionic operators. Schwinger has prescribed  the
insertion of an exponential of a line integral of the gauge field
to make the product gauge invariant. However, his choice was only
one of many possible choices; see, {\it e.g.} \cite{Hag, BGM}. We
have  inserted  an extra factor which involves the field strength
of the gauge field  and  a  nondynamical  function  of  spacetime
coordiantes  and  therefore  does  not  interfere  with the gauge
invariance of the product. This is not  the  most  general  gauge
invariant regularization possible in this approach, but is enough
to   illustrate  the  range  of  possibilities.  By  varying  the
regularization, the equations of motion of  the  Schwinger  model
can be converted to free field equations with the mass exactly as
in  the  usual  case,  or going to zero at {\it both} ends of the
spatial axis or even vanishing everywhere.  In  the  first  case,
there  is  no  fermion in the spectrum at all and the question of
scattering does not arise. In the other two  case,  the  massless
fermion  is  {\it  not} totally reflected, so that the problem of
information loss appears unless further gravitational effects can
change the scenario.

There  is one question which may arise here. Have
we, in changing the definition of the current, changed the model? To be more
specific, the introduction  of  the  $\Phi-$  dependent  term  in
addition  to  just  $A_\mu$  in  the  phase  factor  entering the
point-split current may be suspected to amount to the addition of
an extra interaction.  This  is  not  really  the  case,  as  the
equations  of  motion of the dilatonic Schwinger model itself are
satisfied. The change  is  only  in  the  definition  of  fermion
bilinears as composite operators and this is well known to have a
lot  of  flexibility. Formally, in the limit $\epsilon\to 0$, the
phase factor does reduce to unity, so that the definition of  the
bilinears  adopted in this paper cannot be thought of as changing
the underlying {\it classical} theory. Only the quantum theory,  which
is    not   fully  defined  until  the  definition  of  composite
operators is specified, is  altered. This  alteration  takes  the
form  of  a renormalization of the effective coupling constant in the
theory. The dilatonic field, which entered the model through this
coupling   constant,   can   thus  be  said  to  get  effectively
transformed in the quantum theory. However, this change is not  a
real  one as far as the dilatonic field is concerned. This can be
seen by considering the kinetic energy term of the dilaton field,
which does not get altered. However, in the approximation made by
us following \cite{Strom, PST}, this term is  neglected  and  the
dilaton appears purely as a background field.

Lastly, it should  be  mentioned  that  if  {\it several}  species  of
fermions  are  included,  the problem of information loss appears
automatically \cite{Strom}. This is in keeping with  our  finding
that   magnetically  charged  black  holes  do  not  necessarily behave
like elementary particles in scattering incident fermions.
Therefore, the S-matrices
envisaged by \cite{tH, PST} {\it cannot} always  be  constructed.
The two dimensional model considered here has no horizon, but the
four dimensional model from which it is derived  does  have  one.
There,  the passage of the fermion into the black hole amounts to
a loss of unitarity.

\end{document}